\documentclass[conference]{IEEEtran}
\usepackage{flushend}
\usepackage{multirow}
\usepackage{colortbl}
\usepackage{framed}
\usepackage{amsmath}
\usepackage{subfigure}
\usepackage{threeparttable}
\usepackage{tcolorbox}
\usepackage{indentfirst}
\usepackage{booktabs}

\usepackage[font=small,labelfont=bf]{caption}

\newcommand{\find}[1]{
\begin{tcolorbox}[leftrule=1mm,toprule=0mm,bottomrule=0mm,left=1pt,right=2pt,top=2pt,bottom=2pt]
\em #1
\end{tcolorbox}
}

\def\tablename{Table}
    
\newcommand{\smallsection}[1]{\indent {\bf \underline{#1}}.\hspace{1mm}}
\newcommand{\ea}{{\em et al.}}
\newcommand{\rqone}{What is the impact of temporal inconsistency on the performance of ML-based malware detection approaches?}
\newcommand{\rqtwo}{Why does temporal inconsistency make ML-based malware detection approaches perform so well?}
\newcommand{\rqthree}{How sensitive is the impact of the temporal inconsistency on the accuracy and explanation of the ML-based malware detection approaches?}

\begin{document}

\title{Explainable AI for Android Malware Detection: \\ Towards Understanding Why the Models Perform So Well?}

\author{
\IEEEauthorblockN{
Yue~Liu\IEEEauthorrefmark{2},
Chakkrit~Tantithamthavorn\IEEEauthorrefmark{2}\IEEEauthorrefmark{1},
Li~Li\IEEEauthorrefmark{2}\IEEEauthorrefmark{1}, and
Yepang~Liu\IEEEauthorrefmark{3}}
\IEEEauthorblockA{\IEEEauthorrefmark{2}Faculty of Information Technology, Monash University, Melbourne, Australia. }
\IEEEauthorblockA{\IEEEauthorrefmark{3}Department of Computer Science and Engineering, Southern University of Science and Technology, Shenzhen, China}
Email: \{yue.liu1, chakkrit, li.li\}@monash.edu, liuyp1@sustech.edu.cn

\thanks{*The Corresponding author.}
}

\maketitle

\begingroup\renewcommand\thefootnote{}
\footnotetext{*The Corresponding author.}
\endgroup

\begin{abstract}
Machine learning (ML)-based Android malware detection has been one of the most popular research topics in the mobile security community. 
An increasing number of research studies have demonstrated that machine learning is an effective and promising approach for malware detection, and some works have even claimed that their proposed models could achieve 99\% detection accuracy, leaving little room for further improvement. 
However, numerous prior studies have suggested that unrealistic experimental designs bring substantial biases, resulting in over-optimistic performance in malware detection. 
Unlike previous research that examined the detection performance of ML classifiers to locate the causes, this study employs Explainable AI (XAI) approaches to explore what ML-based models learned during the training process, inspecting and interpreting why ML-based malware classifiers perform so well under unrealistic experimental settings. 
We discover that temporal sample inconsistency in the training dataset brings over-optimistic classification performance (up to 99\% F1 score and accuracy).
Importantly, our results indicate that ML models classify malware based on temporal differences between malware and benign, rather than the actual malicious behaviors.
Our evaluation also confirms the fact that unrealistic experimental designs lead to not only unrealistic detection performance but also poor reliability, posing a significant obstacle to real-world applications. 
These findings suggest that XAI approaches should be used to help practitioners/researchers better understand how do AI/ML models (i.e., malware detection) work---not just focusing on accuracy improvement.

\end{abstract}

\begin{IEEEkeywords}
Android malware detection, Explainable AI, Android security
\end{IEEEkeywords}

\section{Introduction}
\IEEEPARstart{D}{espite} significant and continuous improvements of cybersecurity mechanisms, malware remains one of the most serious threats in cyberspace. 
According to McAfee's report~\cite{mcafeereport2021}, the total number of malware samples reached about 1.5 billion in 2020, with a gradual increase.
Everywhere malware is eroding cyberspace, spawning a slew of subtypes such as mobile malware, MacOS malware, IoT malware, and Coin Miner malware, all of which are causing massive financial losses to both individuals and industries.  

To this end, machine learning (ML) and deep learning (DL)-based malware detection has received significant research attention in recent years.
Specifically, when trained on a large set of data, the built model can distinguish malware from benign samples automatically.
Based on the surveyed results on malware detection by~\cite{qiu2020survey, razgallah2021survey, liu2021deep}, we discovered that ML/DL techniques can generally achieve quite high performance, with detection accuracy reaching up to 99\%, leaving little room for future research.
However, these high-performance approaches appear less viable in practice, as malware defences remain a challenging problem to tackle and malicious applications continue to pose a growing threat to people~\cite{samhi2022difuzer, sun2021characterizing, zungur2021appjitsu, possemato2021preventing, liu2020maddroid}.

Previous studies have reported that existing ML-based malware detection models are susceptible to experimental biases, which could produce unrealistic model performance~\cite{arp2020and}.
Particularly, Pendlebury~\ea~\cite{pendlebury2019tesseract} found that the performance of ML-based malware detection models~\cite{arp2014drebin} is substantially decreased from 90\% to 58\% (F-score) after removing experimental biases (e.g., spatial bias, concept drift).
According to our analysis of the recent research (see Table~\ref{tab:sources_of_temporal_biases})~\cite{qiu2020survey}, we discovered that 29 out of the 30 reviewed relevant studies published between 2014 and 2020 do not consider the realistic experimental design, achieving high accuracy of the ML-based Android malware detection (i.e., over 99\% of F1). 

While accuracy improvement is the primary focus of prior work in ML-based Android malware detection, little research is known about \emph{why the models perform so well}.
Recent research has discovered that the high accuracy of Android malware detection could be due to various sources of experimental biases~\cite{pendlebury2019tesseract}.
However, prior studies still focus only on the predictions (i.e., accuracy) without investigating whether the models can detect malware based on the actual malware-related characteristics or not.
For example, a highly-accurate classifier could correctly classify an image as a dog using background features, which are irrelevant to the dog at all. 
Similarly, ML-based malware detection models could correctly classify an app as malware based on deprecated features (e.g., GET\_TASKS Permission), which may be irrelevant to the actual malicious/benign behaviours.
Therefore, there is a critical need to examine the explanations of the classifications of ML-based Android malware detection as well (i.e., why does the model classify an app as malware?).

\begin{table*}[t]
  \centering
  \scriptsize
  \caption{A list of ML-based Android malware detection that achieves high performance (over 96\% of F1) under an unrealistic experimental setup due to the temporal inconsistency. That means malware samples were chosen from different periods of the benign samples.}
    \begin{tabular}{llccrr}
    \toprule
    \multicolumn{1}{c}{\textbf{Paper}} & \multicolumn{1}{c}{\textbf{Malware sources}} & \textbf{Malware periods} & \textbf{Benign sources} & \multicolumn{1}{c}{\textbf{Benign periods}} & \multicolumn{1}{c}{\textbf{Performance}} \\
    \midrule
    MalDozer~\cite{karbab2018maldozer} & Drebin, Genome, Virushare, Contagio & 2011 - 2017 & Google Play & 2017       & 0.96 F1 score \\
    DeepRefiner~\cite{xu2018deeprefiner} & VirusShare & \textless 2015     & Google Play & 2016       & 0.977 accuracy \\
    Su \ea~\cite{su2016deep} & Drebin, Genome, Contagio & 2011 - 2016 & Google Play & 2016       & 0.995 accuracy, 0.975 F1 score \\
    Khoda \ea~\cite{khoda2019mobile} & Drebin, Genome & 2010 - 2016  & Google Play & 2019       & 0.987 accuracy, 0.985 F1 score \\
    Fan \ea~\cite{fan2020can} & Drebin, Genome, AMD, etc. & 2010 - 2017 & Google Play & 2019 - 2020  & 0.996 precision, 0.977 F1 score \\
    DroidDeep~\cite{su2020droiddeep} & Drebin, Genome, etc. & 2011 - 2017 & Google Play & 2020       & 0.995 accuracy \\
    \bottomrule
    \end{tabular}%
  \label{tab:sources_of_temporal_biases}%
\end{table*}%

\emph{In this paper}, we investigate the impact of the temporal inconsistency on the accuracy and the explanations of the ML-based malware detection.
The \textbf{temporal inconsistency} refers to an unrealistic experimental setup where malware and benign samples are randomly chosen without considering the time period (meaning that malware samples were chosen from 2010, while benign samples were chosen from 2020, see Table~\ref{tab:sources_of_temporal_biases}).
In our experiment, we collected a total of 165,000 Android applications (i.e., 33,000 malware and 132,000 benign applications) that span ten years (2010-2020).
Then, we focus on the three state-of-the-art Explainable ML-based Android malware detection models (i.e., Drebin~\cite{arp2014drebin}, XMal~\cite{wu2020android} and Fan \ea~\cite{fan2020can}).

Our experimental results reveal several important findings:
\begin{itemize}
    \item Temporal inconsistencies between malware and benign in the data significantly increase the detection performance. 
    \item When a temporal inconsistency is introduced in the datasets, the explanations of the ML-based Android malware detection indicate that the models can correctly predict malware based on the temporal-related features, instead of the actual characteristics of malicious and benign behaviors.
    \item Although adjusting the experimental setups like feature sets and malware/benign rates, temporal inconsistencies still unrealistically increase the performance of the ML-based malware detection approaches. 
\end{itemize}

These findings suggest that security analysts should use explainable AI approaches to better understand the models (why the models predict an app as malware or benign?) to better select the most appropriate malware detection models when deciding to deploy them in production.

\textbf{\underline{Novelty.}} To the best of our knowledge, this paper is the first to:

\begin{itemize}
    \item Investigate the impact of temporal inconsistency in Android Malware Detection 
    \item Employ Explainable AI approaches to understand why ML-based malware detection approaches perform so well under temporal inconsistency.
\end{itemize}

\textbf{\underline{Open Science.}} To support the open science initiative, we publish the studied dataset and a replication package, which is publicly available in GitHub.\footnote{https://github.com/yueyueL/XAIforAndroidMalware}

\textbf{\underline{Paper Organization.}}
The remainder of this paper is structured as follows: Section 2 presents the background and related work. Section 3 details our study design. Section 4 provides our experimental results and analysis. Section 5 discusses the study's limitations and potential threats to its validity. Finally, Section 6 concludes the paper.

\section{Background and Related Work}
Researchers raised concerns that many ML-based malware detection techniques are over-optimistic~\cite{arp2020and,pendlebury2019tesseract,li2015potential,zhang2020enhancing}.
In addition, these malware detection approaches were usually black-box models~\cite{liu2021deep}. 
Thus, security analysts often asked questions, e.g., 
How can we trust the predictions of the so-accurate ML-based malware detection models? 
How can we understand whether we are selecting a proper model before deployment? 
To address this challenge, several studies proposed various approaches to explain the predictions of ML-based malware detection models~\cite{arp2014drebin, wu2020android, fan2020can, melis2018explaining, melis2020gradient} (i.e., local explainability).
Below, we introduce our motivational example and summarize the three state-of-the-art explainable ML-based malware detection techniques.

\subsection{Motivation}
Recent research raised concerns that the accuracy of the ML-based malware detection approaches is nearly perfect~\cite{arp2020and,liu2021deep}.
Liu \ea~\cite{liu2021deep} systematically reviewed 132 existing research studies on ML/DL-based Android malware defence approaches. 
Their review results show that most ML-based malware detection approaches achieve an accuracy/F1 measure of 0.98, or even higher.
Moreover, 33 out of 132 surveyed papers present up to 0.99 accuracy/F1 measure, indicating that most ML-based malware detection approaches achieve nearly perfect predictions. 
While existing ML-based malware detection approaches are extremely accurate, it remains unclear why such approaches are so accurate, which still casts some doubt on the research community.

As suggested by prior studies~\cite{pendlebury2019tesseract, zhang2020enhancing, roy2015experimental, miller2016reviewer, cai2020assessing}, the evolution of both the Android platform and Android applications leads to a severe model aging problem (or called time decay, model degradation, and concept drift).
Specifically, malware detection approaches perform poorly on new malware samples.
For example, TESSERACT~\cite{pendlebury2019tesseract} reproduced three state-of-the-art ML-based malware detectors which achieved a high F1 score (up to 0.98), but they found that the performance dropped significantly to 30\% in a time-aware setting (i.e., older apps were used for training and newer ones for testing). 
\textbf{It is still unknown why these Android malware detection approaches perform so well on the original data (i.e., 0.98 F1 score). 
In other words, there is still uncertainty about whether these approaches correctly identify samples based on malware-related characteristics.}

Although the majority of research studies surveyed by~\cite{liu2021deep} did not include information about the time period of collected experimental samples, six relevant primary studies were found, as shown in Table~\ref{tab:sources_of_temporal_biases}. 
It is interesting to observe that these six primary studies collected malware and benign samples from different time periods (i.e., malware samples were older while benign samples were newer). 
This result may be explained by the fact that most malware datasets are usually not maintained or updated after being released, whereas recent benign samples are available via Google Play or third-party markets~\cite{liu2021deep, wang2018android, wang2018beyond}. 
However, prior work~\cite{pendlebury2019tesseract, zhang2020enhancing, roy2015experimental, miller2016reviewer, gao2019understanding, cai2020assessing,liu2021first} has proven that Android malware samples evolve and exhibit distinct characteristics over time. 
As a result, it may lead to unfair predictions, as malware samples and benign samples are collected from distinct time periods.
To the best of our knowledge, no prior literature has studied whether this unfair setting provides a reliable evaluation result for ML/DL-based Android malware detection.
In this study, we use \textbf{temporal inconsistency} to define this problem, which is caused by temporally inconsistent distributions of malware samples and benign samples.

To evaluate the impacts of temporal inconsistency on Android malware detection models, we consider three well-known malware detection approaches using explainable machine learning techniques (i.e., Drebin~\cite{arp2014drebin}, XMal~\cite{wu2020android} and Fan \ea~\cite{fan2020can}). 
First, we would like to stress that we make no specific criticisms of these three approaches.
Because they are available and provide consistent baselines, we choose these three explainable methodologies for our evaluation.

\subsection{Drebin with Linear Support Vector Machine}
Arp~\ea~\cite{arp2014drebin} leveraged an interpretable ML technique, i.e., a linear Support Vector Machine (SVM) to classify if an unknown application is malware or benign. 
To train an SVM-based malware detection model, a linear SVM technique determines a hyperplane that separates both malware and benign classes with maximal margin based on the feature vectors of malware and benign applications in the training data.
To detect the malicious activities of an unknown application, Drebin requires a comprehensive yet lightweight representation of mobile apps.
In particular, Drebin extracts eight feature sets from two main sources.
First, the \emph{manifest} file (i.e., AndroidManifest.xml) is used to store information of the requested hardware components (e.g., camera), the requested permissions (e.g., SEND\_SMS), the list of used Android components (e.g., activities, services, content providers, and broadcast receivers), and filtered intents (e.g., BOOT\_COMPLETED).
Second, the disassembled \emph{dex} code is used to store information of the restricted API calls, used permissions, suspicious API calls, and network addresses.
Then, this information is used to generate a vector representation using a one-hot encoding technique where 1 indicates that an application $x$ contains a feature \emph{x$_i$}, otherwise 0. 
Once the SVM models are trained, the SVM-based malware detection model is applied to classify if an unknown application in testing data is considered as malware or benign.
Finally, Drebin generates an explanation of each prediction using the multiplication ($w_i = w * v_i$) of the feature weights ($w$) of the linear classifier and the actual feature value ($v$) of that test instance ($i$).

\subsection{XMal with Attention Mechanism}
Wu~\ea~\cite{wu2020android} leveraged a multi-layer perceptron (MLP) with the attention mechanism for malware classification, while being able to locally explain the prediction.
Similar to the Debrin~\cite{arp2014drebin}'s approach, the XMal approach leverages feature sets related to API calls and permissions. 
Since there exists a large number of possible features (i.e., 20,000+), the XMal approach selects only the top-154 effective features (including 94 API calls and 60 permissions) for model training.
The XMal approach consists of two layers: the attention layer and the multi-layer perceptron (MLP).
First, a feature vector is generated using a one-hot encoding technique with a dimension of 158.
Then, the feature vector is fed into the attention layer.
The attention layer leverages the attention mechanism proposed by Bahdanau~\ea~\cite{bahdanau2014neural}, which is used to capture the relationship between the features in the input sequence and the next output features, allowing models to retain all the information of the input sequence.
Formally, the attention vector $\alpha_i=(\alpha_i^{(1)}, ..., \alpha_i^{(j)})$ which represents the attention weight of the $j^\mathrm{th}$ feature of the $i^\mathrm{th}$ test instance is computed through a softmax function: $\alpha_{i}^{(j)} = \frac{exp(e_{i}^{(j)})}{\sum_{k=1}^{n}exp(e_{i}^{(k)})}$, where $\alpha_{i}^{(j)}$ denotes the attention weight of the $j^\mathrm{th}$ feature for the $i^\mathrm{th}$ test instance, where $e_{i}^{(k)})$ is the feature vector of the $i^\mathrm{th}$ test instance.
Then, the attention vector is fed into the Multi-layer Perceptron (MLP) layer to map the feature weights into the binary classification.
Finally, the explanation of each prediction is generated based on the attention weights to indicate which features contribute the most to the prediction. 

\subsection{Fan \ea~with Model-Agnostic Explainable Approaches}
Fan \ea~\cite{fan2020can} assessed five different local and model-agnostic explanation approaches (i.g.,LIME~\cite{ribeiro2016should}, Anchor~\cite{ribeiro2018anchors}, LORE~\cite{guidotti2018local}, SHAP~\cite{lundberg2017unified} and LEMNA~\cite{guo2018lemna}) for Android malware analysis.
Unlike model-specific explanation approaches, model-agnostic explanation approaches can explain any machine learning model.
For example, LIME explains a prediction by approximating the decision boundary of any black-box classifier by a simple weighted linear regression model.
Thus, Fan \ea~\cite{fan2020can} evaluated the stability, robustness and effectiveness of model-agnostic explanation approaches on several different malware classifiers (i.e., multilayer perceptron (MLP), random forest (RF), and support vector machines (SVM)).

\subsection{AI/ML-based Experimental Bias}
Researchers have already realized the problem of unrealistic performance in Android malware detection and identified many pitfalls related to high performance, including temporal biases (i.e., time decay or evolved malware) between training and testing data \cite{li2018moonlightbox, liu2022dataset, roy2015experimental,miller2016reviewer, pendlebury2019tesseract, jordaney2017transcend, zhang2020enhancing, xu2019droidevolver, cai2020assessing, yang2021cade}, inappropriate malware rate \cite{allix2016empirical, pendlebury2019tesseract, roy2015experimental, bai2020unsuccessful}, sampling duplication \cite{zhao2021impact}, etc.
These studies have highlighted a correlation between over-optimistic performance and specific unrealistic experimental settings. 
For example, when studying temporal biases, researchers have experimentally found that applications that alter or update over time will cause the trained models to perform poorly on future testing samples \cite{zhang2020enhancing}.
As a result, the actual performance of the proposed ML models might not be as high as the one reported.
Our work takes the initial attempt towards understanding the inner logic of ML-based malware models under unrealistic settings to empirically confirm that learning models could be misled by pitfalls instead of solving the actual task based on benign and malicious behaviors. 

\section{Study Design}

\begin{figure*}[t]
  \centering
  \includegraphics[width=0.9\textwidth]{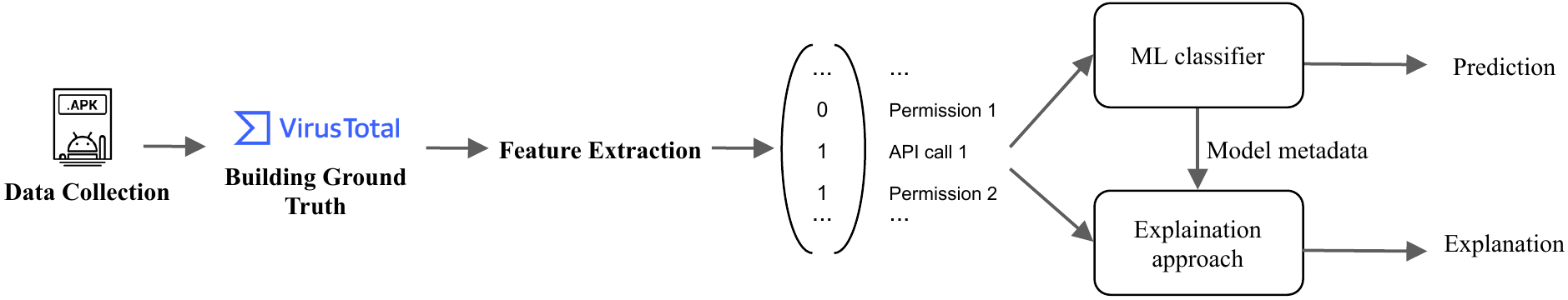}
  \caption{An overview of our experiment: the apps are collected from AndroZoo and labelled by VirusTotal; reverse engineering tools are used to extract features; feature vectors are fed into an ML classifier to generate prediction and explanation. }
  \label{fig:studydesign}
\end{figure*}

\subsection{Goal, Motivation, and Research Questions}

The goal of this paper is to perform a detailed model inspection analysis on the explanations generated by three explainable ML-based malware detection techniques (i.e., Drebin~\cite{arp2014drebin}, XMal~\cite{wu2020android}, and Fan \ea~\cite{fan2020can}).
Such a detailed model inspection analysis could help security analysts better select the most appropriate malware detection models when deciding to deploy in production and help researchers better understand the potential risks associated with unrealistic experimental setups.
To achieve this goal, we aim to address the following three research questions.

\subsection*{(RQ1) \rqone}

\smallsection{Motivation} 
Numerous research studies evaluate malware classification performance using temporally inconsistent datasets, as we discussed before. 
We formulate this research question to ascertain the effect of temporal inconsistency on the performance of machine learning-based malware detection techniques.
Through the replication of three high-profile ML-based malware detection approaches (i.e., Drebin~\cite{arp2014drebin}, XMal~\cite{wu2020android} and Fan \ea~\cite{fan2020can}), we can confirm whether temporal inconsistency results in over-optimistic detection performance.

\subsection*{(RQ2) \rqtwo}

\smallsection{Motivation} 
Only examining classification performance metrics (e.g., accuracy and F1 score) still does not determine what the model is based on to make accurate predictions.
Drebin~\cite{arp2014drebin}, XMal~\cite{wu2020android} and Fan \ea~\cite{fan2020can}~approaches are designed to achieve high accuracy, while being explainable to security analysts.
Therefore, such explainable malware detection techniques allow security analysts to better understand what features contribute to the predictions.
Unfortunately, these three studies have not performed a model inspection analysis on the generated explanations to better understand if the models behave correctly or not.
Such a lack of detailed model inspection analysis for the ML-based malware detection models could lead to inappropriate model selection when deploying them in production (i.e., practitioners still do not know which models to be deployed given the same highly accurate models).
Thus, we formulate this research question to analyze the explanations generated by these three approaches to better understand why such ML-based malware detection approaches under temporal inconsistency are highly accurate.

\subsection*{(RQ3) \rqthree}
\smallsection{Motivation} 
Prior studies~\cite{tantithamthavorn2016icseds,tantithamthavorn2018pitfalls,tantithamthavorn2016comments, arp2020and} raise concerns that the experimental components often have a large impact on the accuracy and explanations of defect prediction models (e.g., data quality~\cite{tantithamthavorn2015icse}, class imbalance~\cite{tantithamthavorn2018impact}, parameter settings~\cite{tantithamthavorn2018optimization}, model validation techniques~\cite{tantithamthavorn2017empirical}).
Similar to ML-based malware detection studies, different studies also use different experimental components~\cite{arp2014drebin, wu2020android, qiu2020survey}.
Yet, little is known about the impact of the experimental components on the accuracy and explanation of explainable malware detection approaches.
As a result, we formulate this research question to gain a better understanding of the impacts of temporal inconsistency under different experimental settings.

\subsection{Experimental setup} 

To address our research questions, our experiment consists of the following steps: (1) data collection; (2) feature extraction; (3) model training \& model evaluation; and (4) model explanations.
Figure~\ref{fig:studydesign} illustrates the overview of our experiment.
We describe each step below.

\subsubsection{Data Collection}
\label{section:dataset}


Generally, ML-based malware detection is formulated as a binary classification task (i.e., classifying whether an application is considered as Malware or Benign).
Thus, it requires samples from two distinct classes (i.e., Malware and Benign apps).
To do so, we download Android applications from the AndroZoo corpus~\cite{allix2016androzoo}.
The AndroZoo corpus consists of a collection of more than 15 million Android applications published between 2010 and 2021, together with the ground-truth labels provided by the VirusTotal software~\cite{pendlebury2019tesseract, cao2020benign, pierazzi2020intriguing}. 
Since the number of applications in the AndroZoo corpus is too large to be studied (15 million), we randomly select a subset of applications in the AndroZoo corpus.

To ensure that our random sample is representative of the population of the AndroZoo corpus, we decide to maintain the same malware ratio as the AndroZoo corpus (i.e., a malware ratio of 17.3\%).
Thus, our studied dataset contains a total of 165,000 Android applications (i.e., 33,000 malware and 132,000 benign applications) that span across a 10-year period (2010-2020).

\subsubsection{Feature Extraction}

Similar to prior studies~\cite{arp2014drebin, wu2020android}, we use the same feature extraction approach to generate feature vectors in order to capture the characteristics of Android applications.
Thus, we use a reverse engineering approach to extract the feature set of Android applications using Androguard,\footnote{https://code.google.com/archive/p/androguard/}.
Androguard is a common open-source tool for static analysis to exploit features like permissions, API calls and activities.

For Drebin, we extract a total of eight feature sets from two main sources, i.e., (1) the \emph{manifest} file (i.e., AndroidManifest.xml), which stores information of the requested hardware components, the requested permissions, the list of used Android components, and filtered intents; and (2) the disassembled \emph{dex} code which stores information of the restricted API calls, used permissions, suspicious API calls, and network addresses.
For Xmal, we extract the same set of 154 features (including 94 API calls and 60 permissions) for model training.\footnote{The original paper claimed 158 features, but they only provided a feature set with 154 features on their personal page.}
For Fan \ea, we use the feature set with XMal since the detailed feature lists are not publicly available.\footnote{The original paper claimed 296 features, including 259 API calls, 22 permissions and 3 intents.}
Then, feature information is used to generate a vector representation using a one-hot encoding technique where one indicates that an application $x$ contains a feature \emph{x$_i$}, otherwise zero.

\subsubsection{Model Training \& Evaluation}

According to Wu~\ea~\cite{wu2020android}, 10-fold cross-validation (CV) is one of the most commonly-used model validation techniques for ML-based malware detection.
Thus, we use a 10-fold CV for model training and model evaluation.
10-fold CV splits a dataset into K partitions, with one partition used for model evaluation and the remaining partitions used for model training.
Then, the process is repeated ten times, and each testing performance is recorded to ensure the stability of the models~\cite{tantithamthavorn2017empirical}. 
For Drebin, we train the model using a linear support vector machine (SVM).
For XMal, we train the model using an attention-based multi-layer perceptron (MLP).
For Fan \ea, we train four different ML models (i.e., MLP, KNN, RF and SVM) with the same settings as those used in the original paper.

\subsubsection{Model Explainability (i.e., Most Important Features)}

Finally, we generate explanations from the Drebin, XMal and Fan \ea~ approaches.
For Drebin with SVM models, we generate an explanation of each prediction using the multiplication ($w_i = w * v_i$) of the feature weights ($w$) of the linear classifier and the actual feature value ($v$) of that test instance ($i$).
For the XMal approach, we generate an explanation of each prediction based on the attention weights to indicate which features contribute the most to the prediction.
For the Fan \ea~approach, we generate an explanation of each prediction based on the LIME approach to indicate which features contribute the most to the prediction.
Note that we don't focus on other model-agnostic explanation approaches discussed in Fan \ea~\cite{fan2020can} since their experimental results demonstrate that LIME provides a better explanation for ML-based malware detection approaches. 
Because we employ 10-fold cross-validation, we record explanations for testing data at each run.

\section{Experimental Results}

In this section, we present the approach and the results of our three research questions.

\begin{table*}[ht]
  \centering  \includegraphics[width=0.8\textwidth]{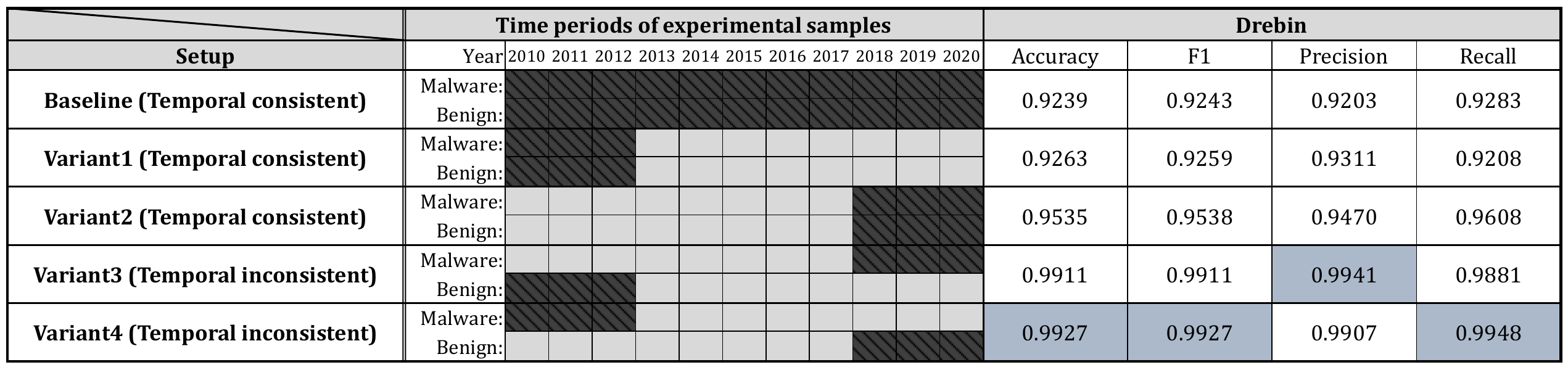}
  \caption{Experimental settings and part of evaluation results of the replication study. The full evaluation results can be found on our online supplementary. The second column (i.e., sample dates) includes 11 years of app samples ranging from 2010 to 2020. The black cell represents that the apps (i.e., all the 3,000 apps for the malware set while randomly selected 3,000 apps from the goodware set)  in the corresponding time frame are selected for training.}
  \label{tab:RQ1motivationperformance}
\end{table*} 

\subsection*{\textbf{RQ1: \rqone}}

In the first research question, we are interested in checking the impacts of temporal inconsistency on detection performance.

\smallsection{Experimental Setup}
In this work, we decide to replicate three prior studies, namely the Drebin, XMal and Fan \ea~approaches, which have been considered the most representative ones available in the community.
For each of these three approaches, we further resort to five training datasets (i.e., settings) to highlight their performances and examine the potential impacts brought by the different settings.
Given a dataset, we first extract the features following the strategies provided by these three approaches, respectively. 
After that, 10-fold cross-validation will be leveraged to assess their performances.
In this work, we evaluate the classification performance through four metrics: Accuracy, F1 measure, Precision, and Recall.


The five settings are detailed as follows.

\begin{itemize}
\item \textbf{Baseline.} For the default setting (hereinafter referred to as the baseline), we select all the malicious apps collected in this work to form the training dataset. As discussed in Section~\ref{section:dataset}, we have prepared 33,000 malicious apps that are released at times ranging from 2010 to 2020, with each year containing 3,000 samples.
To form a balanced training dataset, as highlighted in Table~\ref{tab:RQ1motivationperformance}, we supplement the training dataset by further adding 33,000 benign apps (i.e., with also 3,000 samples per year), which are also randomly selected from the apps collected in this work.

\item \textbf{Variant 1.} The first variant keeps most of the configurations in the Baseline setting except that, in this case, only the samples in the latest three years (i.e., years 2018-2020, as highlighted in Table~\ref{tab:RQ1motivationperformance}) are considered.
In other words, variant 1 forms a balanced training dataset including 9,000 malware and 9,000 goodware.

\item \textbf{Variant 2.} Similar to the setting of Variant 1, in the second variant, the app samples in the first three years (i.e., years 2010-2012, as highlighted in Table~\ref{tab:RQ1motivationperformance}) are considered for training.

\item \textbf{Variant 3.} The readers may have observed that the first two variants (Variants 1-2) have kept the training apps collected from the same period of time. In the third variant, we form the training dataset by collecting the goodware and malware samples from two different periods, i.e., benign samples from the first three years (i.e., years 2010-2012) while malicious samples from the latest three years (i.e., years 2018-2020).

\item \textbf{Variant 4.} The last variant is essentially equivalent to Variant 3 except that, in this case, the benign samples are collected from the latest three years (i.e., years 2018-2020) while the malicious samples are from the first three years (i.e., years 2010-2012).
\end{itemize}

\textbf{Finding 1: The performance of ML-based Android malware detection models could be significantly improved if the temporal inconsistency is introduced in the evaluation dataset.}

Table~\ref{tab:RQ1motivationperformance} summarizes the experimental results for Drebin.
For all the five experimental settings, we are able to achieve high performance with respect to all the four considered evaluation metrics.
The highest case can achieve over 99\% for all four metrics.
These results experimentally confirm that we are indeed able to replicate the high performance of prior studies targeting machine learning-based malware detection.

When comparing the Baseline setting with the four variants, we can observe that the Baseline setting is not able to yield better performance.
Variants 3-4 (temporally inconsistent between malware samples and benign samples) achieve significantly higher performance than the other settings, including that achieved by Baseline and Variants 1-2.
The only difference between Variants 3-4 and others is the involvement of temporal biases, whether the malware and benign datasets are collected from the same time period or not.
This evidence suggests that temporal inconsistency could significantly impact the classifiers' performance if introduced in the experimental datasets of machine learning approaches.
Drebin can achieve the highest detection performance (i.e., 99.27\% accuracy) under Variant 4 where malware samples are older while benign ones are newer.
As for XMal and Fan \ea~, we also observe the similar effects of temporal inconsistencies, which can be found on our online supplementary.
In addition, Table~\ref{tab:sources_of_temporal_biases} and the prior work~\cite{liu2021deep} confirm that Variant 4 is the most common cause of temporal biases between malware and benign samples.
Thus, we mainly focus on discussing the temporal biases that the benign samples are collected from the latest time while the malware samples are collected from the older time in the following experiments.

\find{\textbf{Answer to RQ1:}
Introducing a temporal difference between malware and benign samples in the experimental datasets could significantly increase the detection performance of ML-based Android malware detection approaches.
}

\subsection*{\textbf{RQ2: \rqtwo}}

In the second research question, we are interested in exploring why the aforementioned machine learning approaches can achieve high performances, especially why temporal-related settings (i.e., Variants 3-4) can achieve better performance than non-temporal-related settings (i.e., Baseline and Variants 1-2).
To the best of our knowledge, temporal inconsistency in the training dataset has not been well explored by our fellow researchers yet, and it is still unknown to the community why there is such an impact when temporal inconsistency is introduced in the training dataset.
This question motivates us to go one step further.
To this end, we resort to explainable machine learning techniques to highlight the features that significantly contribute to the classifications, hoping to understand the impact brought by these features with respect to the corresponding training datasets.

\smallsection{Explainable Machine Learning Approach}
After the three ML-based malware detectors output prediction results, an explanation vector \emph{$a_{i}^{k}$} with feature importance values is calculated for each test sample \emph{$x_{i}^{k}$}, where k represents the size of feature list \emph{$S$}. 
For each feature \emph{$S_{j}$}, the average feature importance \emph{$Avg\_fi(S_{j})$} can be calculated as: $Avg\_fi(S_{j}) =  \frac{1}{N} \cdot \sum_{i =1}^{N} a_{i}^{j}$, where N is the size of test samples. 
Thus, we can obtain the average feature importance for each characteristic when the model generates the predictions on a test set. 

Except calculating average feature important, we sort the feature importance vector \emph{$a_{i}^{k}$} and count the frequency \emph{$Count\_top(S_{j})$} whether characteristic exist in top features:
$Count\_top(S_{j}, T) =  \frac{1}{N} \cdot \sum_{i =1}^{N} [S_{j}\in top(a_{i}^{k},T)]$, 
where function \emph{$top(a_{i}^{k},T)$} means getting the top \emph{T} important features from the feature importance vector \emph{$a_{i}^{k}$}.
Then, this equation will judge whether feature \emph{$S_{j}$} exists in top T important features of \emph{$x_{i}^{k}$}. 
A proportion of feature \emph{$S_{j}$} in top features would be calculated.

\begin{table*}[t]
  \centering
  \caption{Examples of time-specific features}
  \includegraphics[width=0.95\textwidth]{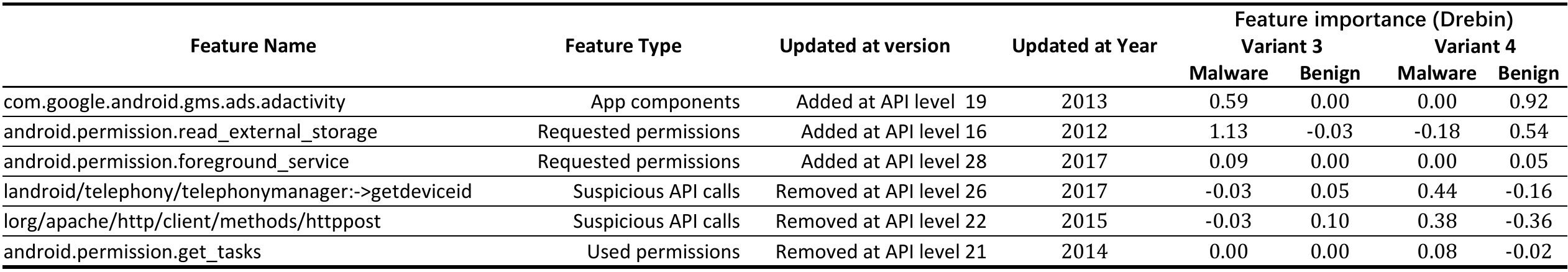}
  \label{tab:time-specific_features_example}%
\end{table*}%

\smallsection{Experimental Setup}
To help readers better understand this work, we resort to the same experimental settings proposed for answering RQ1 to fulfill the experiments of RQ2, i.e., three approaches with five settings constructed with balanced training datasets.
The only difference is that, when re-running the machine learning classifications, we apply the explanation module mentioned above to the original classification so as to further collect the feature importance information for each testing example.
Prior studies~\cite{melis2018explaining, arp2014drebin} have proven that the selection of feature sets plays an important role in explaining machine learning-based malware classifications, for which their performances are often decided by a small number of top-ranked features.
Thus, our follow-up detailed analyses hence mainly focus on the top-ranked features.

\textbf{Finding 2: When applied to ML-based malware detection, the top-ranked features highlighted via explainable machine learning approaches may not always capture the difference between malicious and benign behaviors. They could simply be time-specific features that only exist in either historical or latest apps.}

As discussed previously, there is a strong correlation between the evaluation performance of ML classifiers and the temporal distribution of the training samples.
In this RQ, we hence resort to explanations of ML classifiers for each testing sample to determine if such temporal distributions will impact the classification results.
Specifically, in this work, we have identified two types of time-specific features: (1) Added ones and (2) Removed ones, which are respectively defined as follows.

\begin{itemize}
\item \textbf{Added Features.}
Features that are added to the Android framework after the apps are released to the ecosystem. 
Therefore, these apps, either malicious or benign, will have no chance to access those features. 
However, the remaining apps, which are released after the time when the features are added, may have the opportunity to include those features. 
It hence introduces biases specific to time rather than maliciousness.
Table~\ref{tab:time-specific_features_example} includes several added features on Drebin's predictions.
For example, the "gms.ads.adactivity" app component was only added at Android 4.4 (or API level 19, the first revision released in 2013) to allow apps to display advertisements and earn revenue.
Google Mobile Ads APIs became a part of Google Play services (gms) in Oct. 2013, so the apps released before 2012 will have unlikely included this characteristic, while apps released after 2018 could have.
Under Variant 4 where malware samples are before 2012 while benign samples are after 2018, Drebin considers this ads-related feature as one of the top features in terms of benign identification, with 0.92 feature importance, while a zero value for identifying malware.
Yet, when Drebin is trained under Variant 3 with malware data after 2018 but benign data before 2012, this feature is recognized as a malware-related feature, with 0.59 feature importance but performs a low impact on benign identification. 
From Table~\ref{tab:time-specific_features_example}, added features usually have higher feature importance in identifying the category containing the latest samples, whether the category is benign or malware.

\begin{table}[htbp]
  \centering
  \scriptsize
  \caption{Comparison of feature importance of time-specific features for malware/benign prediction by Drebin. The chart records the ratios of test samples containing relevant time-specific features in top features when ML classifiers make predictions. The ground truth of temporal information of each feature is generated based on the official Android Developer Documentation\protect\footnotemark.}
    \begin{tabular}{l|r|c|c|c|c}
    \toprule
    \multicolumn{1}{r}{} &       & \multicolumn{2}{c|}{Top 10} & \multicolumn{2}{c}{Top 20} \\
    \midrule
    \multicolumn{1}{c}{} &       & Added & Removed & Added & Removed \\
    \midrule
    \multirow{2}[2]{*}{\textbf{Baseline}} & Malware & 0.3342 & 0.8303 & 0.5230 & 0.8640 \\
          & Benign & 0.4935 & 0.3705 & 0.5783 & 0.5765 \\
    \midrule
    \multirow{2}[2]{*}{\textbf{Variant 1}} & Malware & 0.1402 & 0.9010 & 0.2163 & 0.9018 \\
          & Benign & 0.1097 & 0.3727 & 0.1229 & 0.4093 \\
    \midrule
    \multirow{2}[2]{*}{\textbf{Variant 2}} & Malware & 0.4818 & 0.3535 & 0.8194 & 0.5752 \\
          & Benign & 0.6697 & 0.1575 & 0.8216 & 0.3941 \\
    \midrule
    \multirow{2}[2]{*}{\textbf{Variant 3}} & Malware & 0.9259 & 0.3015 & 0.9342 & 0.6941 \\
          & Benign & 0.0968 & 0.4312 & 0.1142 & 0.4871 \\
    \midrule
    \multirow{2}[2]{*}{\textbf{Variant 4}} & Malware & 0.1834 & 0.8445 & 0.1970 & 0.8834 \\
          & Benign & 0.9047 & 0.2503 & 0.9187 & 0.6471 \\
    \bottomrule
    \end{tabular}%
  \label{tab:RQ2top10}%
\end{table}%
\footnotetext{https://developer.android.com/reference}

\item \textbf{Removed Features.}
Features that have been deleted (or deprecated) from the Android framework at some stage, and hence the apps released after that will unlikely access them.
In this work, we also consider deprecated ones as removed. 
Although deprecated features are still available, they are explicitly discouraged from being used anymore. 
Likely, developers who follow the official recommendations will no longer use them.
Table~\ref{tab:time-specific_features_example} also includes several removed features on Drebin's predictions.
For example, the "TelephonyManager.getDeviceId" API call were deprecated from Android 8.0 (or API level 26, first revision released in 2017), as this new release updated new API calls to return the unique device ID. 
With the same conclusion with \cite{arp2014drebin, wu2020android}, the key features relevant to malware identification outputted by Drebin under Variant 4 include "getDeviceID", with 0.43 and 0.13 feature importance respectively.
However, when Drebin are trained under Variant 3 with malware data after 2018 but benign before 2012, this risky feature is recognized as a benign-related feature.
From Table~\ref{tab:time-specific_features_example}, removed features usually have higher feature importance to identify the category containing historical samples, whether the category is benign or malware.
\end{itemize}

\begin{table*}[t]
  \centering
  \scriptsize
  \caption{Results of prediction performance and feature importance of time specific features when ML-based malware classifier is trained on one temporally
  inconsistent dataset and is tested on another one (i.e., Variant 3 and Variant 4). }
    \begin{tabular}{lccccrcccc}
    \toprule
               & \multicolumn{4}{c}{\multirow{2}[4]{*}{\textbf{Prediction Performance}}} & \multicolumn{5}{c}{\textbf{Explanation results}} \\
\cmidrule{6-10}               & \multicolumn{4}{c}{}                              &            & \multicolumn{2}{c}{\textbf{Top 10}} & \multicolumn{2}{c}{\textbf{Top 20}} \\
\cmidrule{2-10}               & \textbf{Accuracy} & \textbf{F1} & \textbf{Precision} & \textbf{Recall} &            & \textbf{Added} & \textbf{Removed} & \textbf{Added} & \textbf{Removed} \\
    \midrule
    \multirow{2}[1]{*}{\textbf{Trained on Variant 3, tested on Variant 4}} & \multirow{2}[1]{*}{0.0968} & \multirow{2}[1]{*}{0.0368} & \multirow{2}[1]{*}{0.0395} & \multirow{2}[1]{*}{0.0346} & Malware    & 0.8754     & 0.7282     & 0.9074     & 0.8660 \\
               &            &            &            &            & Benign     & 0.1151     & 0.4864     & 0.1366     & 0.8033 \\
    \multirow{2}[1]{*}{\textbf{Trained on Variant 4, tested on Variant 3}} & \multirow{2}[1]{*}{0.1415} & \multirow{2}[1]{*}{0.0536} & \multirow{2}[1]{*}{0.0598} & \multirow{2}[1]{*}{0.0487} & Malware    & 0.0831     & 0.5626     & 0.0965     & 0.5695 \\
               &            &            &            &            & Benign     & 0.7970     & 0.0778     & 0.8010     & 0.2759 \\
    \bottomrule
    \end{tabular}%
  \label{tab:RQ2_reverse_test}%
\end{table*}%

Table~\ref{tab:RQ2top10} summarizes the ratio of time-specific features involved in each classification of the five experimental settings by Drebin.
Among top $k$ important features observed, the ratio of added and removed features (i.e., say $x$ and $y$) are calculated via $x/k$ and $y/k$, respectively.
Since only a small number of features will be regarded as important ones, as we experimentally discovered previously, in this table, we only summarize the ratio based on top-10 and top-20 features ranked based on their importance.

  

As indicated in the Baseline setting, the ratios of added and removed features are quite high. This is expected as this setting has included a wide range of samples (i.e., from 2010 to 2020).
The ratios of added features, w.r.t. predicting malware and benign apps, are more or less the same, as the ratios of removed features have a slight discrepancy.
Similar results could also be observed in Variants 1\&2 settings as both of them have collected app samples (i.e., both malware and benign apps) from the same period.
However, when comparing the results obtained in Variants 3\&4, for which the malware and benign samples are collected from different time periods, we could observe clear differences. 
Under Variant 3, we observe that when Drebin makes decisions to identify malware, 92\% of malware in testing samples contain newly added features in the top 10 features, but only 9\% benign in testing samples contain added features in the top 10 features. 
The observation is not surprising since the malware samples in this setting are collected from apps released from 2018 to 2020 but benign from 2010 to 2012. 
On the contrary, when looking at Variant 4, where malware samples are from 2010 to 2012, but benign samples are created after 2018, Drebin can build distinguish rules indicating that the benign identification greatly depends on newly added features while malware identification highly depends on removed (or deprecated) characteristics. 
This evidence indicates that the important features contributing to the high performance of Drebin may not necessarily be related to apps' maliciousness (or benignness) but could simply be discrepancies introduced by temporal inconsistencies in the training dataset.
When experimental data is temporally inconsistent, newly added features have a higher positive impact on identifying the category collected on a later date, while deprecated/removed features have a higher positive impact on the category collected on an earlier date.

\textbf{Finding 3: When using testing samples from distinct periods, ML models still distinguish malware/benign based on temporal differences learned from training data, resulting in extremely poor performance. }

To further understand the impacts of temporal inconsistency, we obtain the models from RQ1 with the best performance under Variant 3 and Variant 4, respectively, and test the performance on another temporally inconsistent dataset.
Table~\ref{tab:RQ2_reverse_test} presents the prediction performance and relevant explanation results. 
When Drebin is trained on Variant 4 (malware is from 2010-2012 while benign is from 2018-2020), but tested on Variant 3 (malware is from 2018-2020 while benign is from 2010-2012), it only obtains 14\% accuracy, which is much less than 99\% obtained in RQ1.
The explanation results show that the trained model under Variant 4 still thinks that the samples with more added features are more likely to be benign, while the samples with more removed features are more likely to be malware. 
Similarly, when Drebin is trained on Variant 3 but tested on Variant 4, the results show that Drebin considers the samples with more added features as malware, causing only 9\% accuracy.
Therefore, this experiment demonstrates that if the time difference between malware and benign changes, the ML-based model trained under temporal biases doesn't work.
This observation further demonstrates that ML-based malware detectors distinguish malware from benign based on time-specific features under the temporal inconsistency.

\begin{table}[htbp]
\scriptsize
  \centering
  \caption{Evaluation results of the replication study under temporal biases (Malware is from 2010-2012 while benign is from 2018-2020)}
    \begin{tabular}{lccrcc}
    \toprule
               & \multicolumn{2}{c}{\textbf{Prediction Performance}} & \multicolumn{3}{c}{\textbf{Explanation Results}} \\
               & \textbf{Accuracy} & \textbf{F1} &            & \textbf{Added} & \textbf{Removed} \\
    \midrule
    \multirow{2}[1]{*}{\textbf{Drebin}} & \multirow{2}[1]{*}{0.9927} & \multirow{2}[1]{*}{0.9927} & Malware    & 0.1834     & 0.8445 \\
               &            &            & Benign     & 0.9047     & 0.2503 \\
    \multirow{2}[0]{*}{\textbf{Xmal}} & \multirow{2}[0]{*}{0.9822} & \multirow{2}[0]{*}{0.9824} & Malware    & 0.0104     & 0.8180 \\
               &            &            & Benign     & 0.7702     & 0.7723 \\
    \multirow{2}[0]{*}{\textbf{MLP + LIME}} & \multirow{2}[0]{*}{0.9861} & \multirow{2}[0]{*}{0.9861} & Malware    & \multicolumn{1}{r}{0.00821} & 0.7790 \\
               &            &            & Benign     & \multicolumn{1}{r}{0.81651} & 0.7704 \\
    \multirow{2}[0]{*}{\textbf{RF + LIME}} & \multirow{2}[0]{*}{0.9869} & \multirow{2}[0]{*}{0.9869} & Malware    & \multicolumn{1}{r}{0.00687} & 0.7819 \\
               &            &            & Benign     & \multicolumn{1}{r}{0.81603} & 0.7670 \\
    \multirow{2}[1]{*}{\textbf{SVM + LIME}} & \multirow{2}[1]{*}{0.9806} & \multirow{2}[1]{*}{0.9807} & Malware    & 0.0238     & 0.7709 \\
               &            &            & Benign     & 0.8274     & 0.7806 \\
    \bottomrule
    \end{tabular}%
  \label{tab:rq2_all_explain}%
\end{table}%

\textbf{Finding 4:  If temporal inconsistency exists, all three ML-based malware detection approaches provide highly accurate predictions based on temporal differences.}

Table~\ref{tab:rq2_all_explain} presents the prediction results and explanation results of three ML/DL-based malware detection approaches under Variant 4.
We can observe that six ML-based Android malware detection approaches achieve fairly high performance, where all the accuracy and F1 score are higher than 0.98 under temporal inconsistency.
This finding indicates that temporal biases between the experimental malware and benign samples influence predictions regardless of the type of machine learning algorithms.
From the explanation results, we can observe that ML-based Android malware detection approaches capture time differences between malware and benign (i.e., under Variant 4, malware is likely to include removed features while benign is likely to include added features).

\begin{figure*}[ht]
  \centering
  \includegraphics[width=0.8\textwidth]{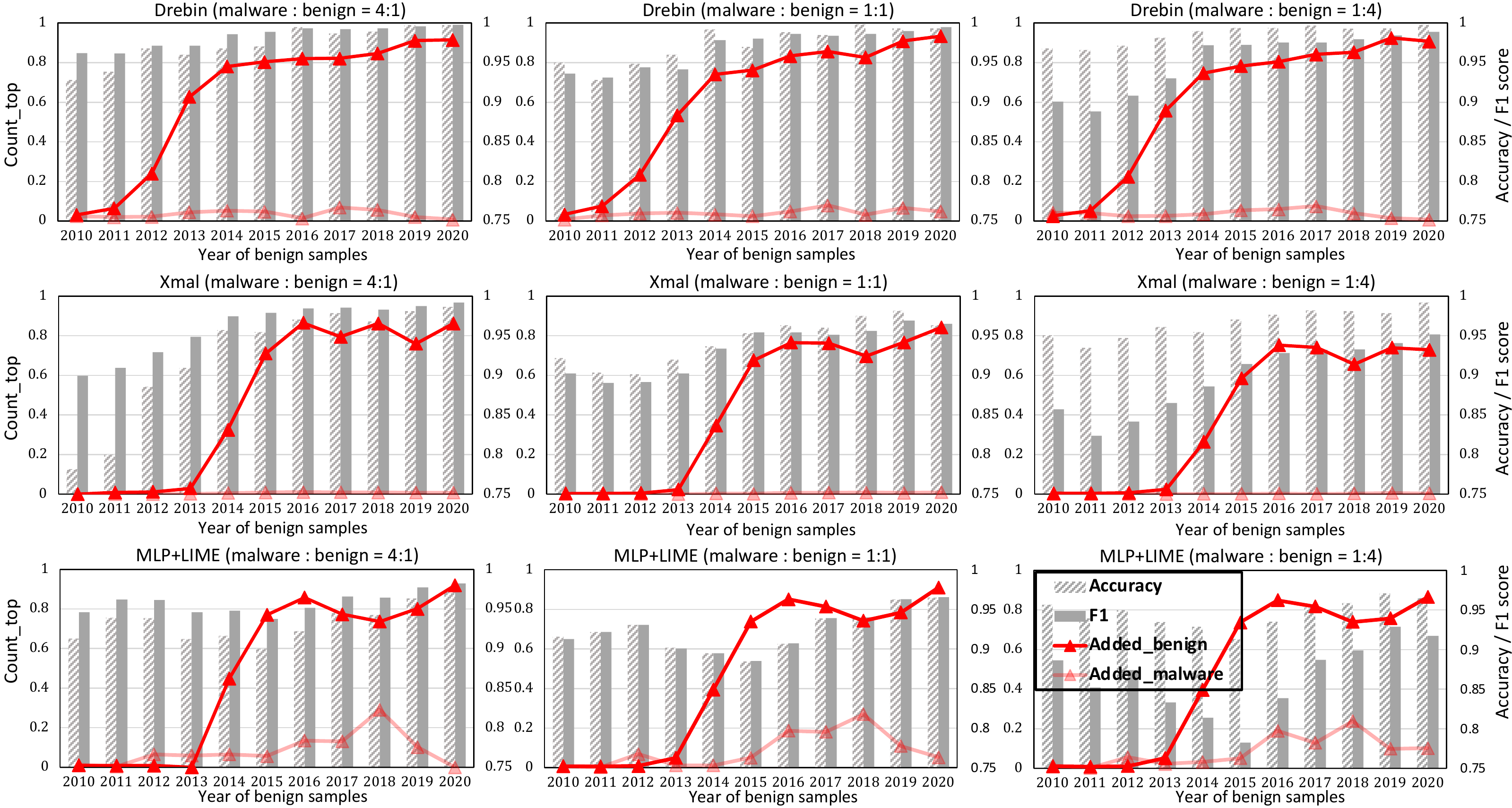}
  \caption{Experimental results obtained via Setting 1 that keeps the malware data untouched (in  the year 2010) and train the models on different benign samples collected at different years (from 2010 to 2020)}
  \label{fig:1.3added_features}
\end{figure*}

\find{\textbf{Answer to RQ2:}
Many time-specific features are only available in either historical or latest applications.
Explanations for testing samples reveal that ML models correctly identify the temporal differences between malware and benign samples, resulting in high performance. 
}

\subsection*{\textbf{RQ3: \rqthree}}

Our previous experiments have empirically demonstrated that the performance of ML-based malware classification approaches could be significantly increased when temporal inconsistencies are introduced in the training dataset.
Especially on Variant 4, malware samples are from an earlier time than benign samples, achieving up to 98\% accuracy for all ML-based malware classifiers. 
Actually, Variant 4 is commonly occurring in the research domain, as shown in Table~\ref{tab:sources_of_temporal_biases}.
The finding is however only confirmed through a limited number of experimental settings, letting it unknown if it holds true for other experimental settings (i.e., the finding per se is generic).
To this end, in the last research question, we would like to explore this by conducting large-scale experiments with different settings.
Specifically, we confirm the genericity through varying settings with customized malware/goodware ratios and temporal sample inconsistencies.

\smallsection{Experimental Setup}
The purpose of this section is used to further confirm the correlation between inconsistent temporal distributions and classification performances when the rate of malware to benign is changed. 
To this end, we split our obtained dataset obtained in Section \ref{section:dataset} into 22 subsets based on the APK type (Malicious/benign) and appearance time (2010 to 2020). 
Specifically, to investigate the influence of time interval size on final prediction performance, we consider the following time-related settings: keep the malware data untouched (in the year 2010) and train the models on different benign samples collected in different years
Except that, we further consider the impacts over three malware/benign ratios: a balanced dataset with malware/benign ratio (i.e., 1:1), a large malware set with malware/benign ratio (i.e., 4:1), and a smaller malware set with malware/benign ratio (i.e., 1:4).

\textbf{Finding 5: Varying the Malware/Benign Rates in the training dataset will have a great impact on the performance of machine learning-based malware detection approaches.}

Figure \ref{fig:1.3added_features} further present the performance results and the proportion of time-related features in top 10 features (\emph{$Count\_top(S_{j}, 10)$}) obtained via the explanable AI approach.
What stands out in the figures is that malware rates can influence detection performance.
When malware/benign rate is set to 4:1, three ML-based malware classifiers always present a much higher F1 score than the other two at all time points.
When the malware/benign rate is set to 1:4, the malware detectors obtain the lowest F1 score.
These results mirror those of the previous studies~\cite{pendlebury2019tesseract, allix2016empirical} that have examined the impacts of unrealistic malware rates on ML-based malware classifiers. 
As Pendlebury et al.~\cite{pendlebury2019tesseract} described, most mobile applications in the real world are benign samples, but most research studies build an unrealistic malware rate, causing over-optimistic detection performance.

\textbf{Finding 6: When temporal inconsistencies between malware and benign get bigger (with a larger time interval size), ML classifiers tend to achieve ``better'' performance.}

From Figure~\ref{fig:1.3added_features}, it can be seen that Drebin, XMal and Fan \ea~can often achieve higher F1 and accuracy values regardless of malware rate, when temporal biases between malware and benign become larger.
The first two subfigures of Figure~\ref{fig:1.3added_features} show that when the time interval between malware in 2010 and benign gets bigger, the detection performance of ML-based malware classifiers gradually improves. 
Indeed, there is a steady increase in the proportion of benign applications with newly added features in their top 10 key features as time increases when 2010 malware data is combined with variable benign data at different time spots.
This observation indicates that as benign samples evolve, added features become increasingly important in distinguishing benign from historical malware samples. 
The experimental results further support our previous finding that ML classifiers learn the temporal differences between malware and benign samples, resulting in unrealistic performance to rise.

Overall, our observations suggest that the rules for distinguishing malware built by the ML models strongly depend on the temporal distribution of the training malware and benign samples.
When training data is inconsistent in time, malware identification of the ML-based approaches is highly reliant on learning temporal differences, and the temporal differences are reflected in a wide range of characteristics. 
Explanations for testing samples reveal that the feature importance of time-specific features gradually increases as the unrealistic performance improves. 
Further analysis of key feature explanations reveals that temporal differences are related to a wide range of features in the feature sets.

\find{\textbf{Answer to RQ3:}
The positive correlation between temporal sample inconsistency in the training dataset (regardless of balanced or imbalanced malware/benign sample sets) and the ML-based classification results is generic.
When the temporal inconsistencies between malware and benign samples are greater, ML classifiers learn a greater number of time-related differences, which subsequently contribute to higher prediction performances. 
}

\section{Discussion}
\label{sec: discussion}


\textbf{Explainability of Malware Detection.} 
In this study, we explore three explainable machine learning-based malware classifiers.
Currently, more complex deep learning algorithms, such as recurrent neural networks and conventional neural networks, are becoming more popular for building malware classifiers because they could provide better detection performance without a feature selection process.
However, these algorithms are usually black-box models with limited explainability.
Although prior works suggest that complex deep neural networks boost performance, our experiments have demonstrated that even simple ML models can achieve a high performance up to 99\% accuracy when experimental samples are temporally inconsistent. 
We explore the over-optimistic and unreliable experimental results caused by unrealistic evaluation designs, which have no direct connection with the types of classifiers used.
Four types of ML models (i.e., SVM, attention-based neural network, MLP and RF) generate consistent evaluation results, confirming the generality of our findings for other machine learning-based malware detection approaches.
In addition, we investigate two types of explainable AI approaches including model-specific explainable approaches (i.e., linear SVM and attention-based neural networks) and model-agnostic explainable approaches (i.e., LIME), confirming the validity of explainable AI approaches for analyzing or improving ML-based Android malware detection approaches.

\textbf{Time-specific Features.}
We define the time-specific features based on the official descriptions of the Google developer documentation.
By inspecting the feature importance of time-specific features, we found that ML-based malware detection approaches learn temporal differences to identify malware from benign when training data is temporally inconsistent.
When using a smaller feature set with fewer time-specific features, the explanation results of XMal and Fan \ea~are not always consistent with that of Drebin in RQ2, but from RQ3, we found the performance of XMal and Fan \ea~ is highly correlated with time biases in the training data. 
The explanation is that except for the time-specified features we defined based on Google Developer Documentation, temporal differences depend on much more complex factors. 
For example, we output the top 10 features for benign identification when Drebin is trained on 2010 malware and 2020 benign of RQ3.
We observe that three added features are regarded as key benign-related features but these features have no impact when benign is from the historical time. 
What is surprising is that other top 10 features also only show a high feature importance only when benign is from the latest period.
This result can be explained by the fact that the temporal differences are represented not just in the added features we defined based on the Google Developer Documentation. 
Our motivation is not to locate all time-specific features, but we utilize the feature importance of these features to determine whether malware detectors are reliable. 
More importantly, the evaluation results help us confirm that temporal biases can't be eliminated by feature selection or reduction.
Although XMal and Fan \ea~ only use 154 features, our evaluation results of RQ3 show that their detection performance is highly related to temporal inconsistency.

\textbf{Threats to Validity}
The primary threat to internal validity lies in the implementation of the study. 
To reduce this threat, we utilized three ML-based malware detection approaches.  
The external threat to validity mainly lies in the used datasets. 
We collect malware and benign samples from the Androzoo repository, which comprises a collection of more than 15 million Android samples from various application markets. 
We follow the same process with the reproduction study to process the application and construct the feature vectors. 
To further investigate the generality of our findings, we evaluate the three approaches on the different period data.

\section{Conclusion}
The paper utilized explainable malware detection models to investigate why the existing research works present highly accurate performance. 
By evaluating the explanation results of ML models, we found that most of the results are not realistic since ML models haven't figured out the real difference between malware and benign. 
Specifically, accurate predictions are strongly related to the temporal inconsistency in training data. 
Our work demonstrates that a robust experimental setup for malware classification models is required, otherwise, ML/DL-based models present over-optimistic results.
We encourage the community to jump outside of the ideal world of high-performance of machine learning and should focus more on reliability and applicability, not only on classification evaluation metrics. 
Furthermore, explainable AI techniques helped us understand the inner logic and infer the decision reasons for ML-based Android malware detection models.
We expect that our work will inspire future researchers to utilize explainable AI techniques to explore the underlying issues in ML/DL-based systems.

\section*{acknowledgment} 

The authors would like to thank the anonymous reviewers who have provided insightful and constructive comments on this paper. 
Chakkrit Tantithamthavorn was partly supported by the Australian Research Council's Discovery Early Career Researcher Award (DECRA) funding scheme (DE200100941). 
Li Li was partly supported by the Australian Research Council
(ARC) under a Discovery Early Career Researcher Award (DECRA) project DE200100016, and a Discovery project DP20010002.

\bibliographystyle{IEEEtran}
\bibliography{reference}

\end{document}